\documentclass[11pt]{article}
\usepackage{varioref,exscale,latexsym,amsmath,amssymb}
\usepackage[dvips]{graphicx}
\title{Quantum Gravitational Corrections to the Nonrelativistic
Scattering Potential of Two Masses}
\author{N.~E.~J Bjerrum-Bohr$^a$\\
John F. Donoghue$^b$, and Barry R. Holstein$^{b}$\\
$^a$ Department of Physics\\
University of California at Los Angeles\\
Los Angeles, CA 90095-1574, U.S.A\footnote{\footnotesize{Address
until Dec. 20th}}\\
and \\
The Niels Bohr Institute,\\
Blegdamsvej 17, Copenhagen \O, \\
DK-2100, Denmark\\
$^b$ Department of Physics-LGRT\\
University of Massachusetts\\
Amherst, MA  01003}
\begin{document}
\begin{titlepage}
\maketitle
\end{titlepage}
\begin{abstract}
We treat general relativity as an effective field theory, obtaining the
full
nonanalytic component of the scattering matrix potential to one-loop
order.
The
lowest order vertex rules for the resulting effective field theory are
presented and
the one-loop diagrams which yield the leading nonrelativistic
post-Newtonian
and
quantum corrections to the gravitational scattering amplitude to second
order
in $G$
are calculated in detail. The Fourier transformed amplitudes yield a
nonrelativistic
potential and our result is discussed in relation to previous
calculations.
The
definition of a potential is discussed as well and we show how the ambiguity
of
the
potential under coordinate changes is resolved.

\end{abstract}
\maketitle
\section{Introduction}
The idea that a field theory need not be strictly renormalizable
in the traditional sense yet can still yield useful quantum
predictions when treated as an effective field
theory~\cite{Weinberg:1978kz} has been clearly demonstrated in
chiral perturbation theory and in other applications\cite{gdh}.
Quantum loop calculations lead to well defined results in the low
energy limit. Interestingly, such methods can also be applied to
general relativity. As an effective field theory, the quantization
of general relativity can be carried out in a consistent way,
since the troublesome singularities which occur for various types
of matter sources in traditional renormalization
schemes~\cite{Veltman1,Veltman2,Goroff:1985th,Deser:cz, Deser:cy}
can be absorbed into phenomenological constants which characterize
the effective action of the theory. This effective field theory
approach offers then a possible way around the familiar
renormalization difficulties of general relativity in the low
energy regime and, using this approach with background field
quantization~\cite{Dewitt} of general relativity, one of
us~\cite{Donoghue:1993eb, Donoghue:dn} some years ago derived the
leading quantum and classical corrections to the Newtonian
potential of two large non-relativistic masses. This calculation
has since been the focus of a number of
publications~\cite{Muzinich:1995uj,Hamber:1995cq,akh,Thesis}, and
this work continues, most recently in the paper~\cite{ibk}.
Unfortunately, due to difficulty of the calculation and its myriad
of tensor indices there has been little agreement among these
various authors. The classical component of the correction has
previously been discussed in the papers
by\cite{Gupta,Iwasaki,Okamura,Barker:bx,Barker:ae}, and here there
is general agreement although, as we shall discuss, there exists
an unavoidable ambiguity in defining the potential.  The basic
disagreements lie rather in the quantum corrections and in the
present paper we shall present what we believe to be the {\it
definitive} result for the leading classical and quantum
corrections of order $G^2$, using the full scattering amplitude as
the definition of the non-relativistic potential.

We note that, as a prelude to this effort, in a recent
paper~\cite{Donoghue:2001qc} two of us have dealt with the the
quantum and classical corrections to the Reissner-Nordstr\"{o}m
and Kerr-Newman metrics of charged scalars and fermions. Such
quantum and classical corrections have also been considered from
the viewpoint of a scattering potential in a paper~\cite{emg} by
one of us. Recently we have also calculated the full classical and
quantum corrections to the Schwarzshild and Kerr metrics of
scalars and fermions~\cite{metric} and have shown in detail how
the higher order gravitational contributions to these metrics
emerge from loop calculations. In the present paper then we
consider the corresponding calculation of the full scattering
amplitude.

Of course, treating general relativity as an effective field
theory is carried out at the cost of introducing a never ending
set of additional higher derivative couplings into the theory. In
this sense Einstein's general relativity is still a perfectly
valid theory for gravitational interactions---although now it
represents only the the {\it minimal} theory. At some stage
additional derivative couplings must be appended to the Einstein
action, signifying manifestations of the higher energy component
of the effective field theory. However, the low energy scattering
potential is free from these new couplings and represents a
model-independent result for quantum gravity.

This calculation is possible because the post-Newtonian and quantum
corrections
which we consider are determined fully by the {\it non-analytic} pieces of
the one
loop amplitude generated by the lowest order Einstein action. Of course,
in order to
deal with the ultraviolet divergences which arise at one loop, one must
renormalize
the parameters of higher derivative terms in the action. However, such
pieces will
only affect the analytic parts of the one-loop amplitude, and will not
contribute to
our potential.

We will employ the same conventions as in our previous papers, namely
($\hbar = c =
1$) as well as the Minkowski metric convention $(+1,-1,-1,-1)$. We will
begin in
section 2 with a very short introduction to the effective field theory
quantization
of general relativity and focus here especially on the distinction between
non-analytic and analytic contributions to the scattering amplitude.  We
also
include a discussion of the definition of the non-relativistic potential.

Next in section 3 we evaluate the diagrams which contribute to the
scattering
and
examine in detail the results for the various components. The resulting
nonanalytic
piece of the scattering amplitude is then used in order to construct the
leading
corrections to the nonrelativistic gravitational potential. We also
discuss
our
result in relation to previous calculations and attempt to sort out the
various
inconsistencies in the published numbers. Finally in a concluding section
we
summarize our findings.

\section{Review of general relativity as an effective field theory}
We begin with a brief review of general relativity, the Lagrangian of
which (not
including a cosmological term) is
\begin{equation}
{\cal L} = \sqrt{-g}\left[\frac{2R}{\kappa^2}+ {\cal
L}_{\text{matter}}\right]\label{eq:lgn}
\end{equation}
where $\kappa^2 = 32\pi G$ is the gravitational coupling,
$R^\mu_{\ \nu\alpha\beta} \equiv
\partial_\alpha \Gamma_{\nu\beta}^\mu
-\partial_\beta \Gamma_{\nu\alpha}^\mu
+\Gamma^\mu_{\sigma\alpha}\Gamma_{\nu\beta}^\sigma
-\Gamma_{\sigma\beta}^\mu\Gamma_{\nu\alpha}^\sigma$ is the curvature
tensor, and $g$
denotes the determinant of the metric field $g_{\mu\nu}$. Here
$\sqrt{-g}{\cal
L}_{\rm matter}$ is a covariant Lagrangian for the matter fields, and in
principle
any type of matter field could be included. This action defines the
classical theory
of general relativity.

In order to treat Eq. \ref{eq:lgn} as an {\it effective} field
theory one must include all possible higher derivative couplings
of the fields in the gravitational Lagrangian. In this way any
field singularities generated by loop diagrams can be associated
with some component of the action and hence can be absorbed via a
simple redefinition of the coupling constants of the theory.
Treating all such coupling coefficients as experimentally
determined quantities, the effective field theory is then finite
and contains no singularities at any finite order of the loop
expansion.

We can consequently write an effective action for pure general relativity
as
\begin{equation}
{\cal L} =
\sqrt{-g}\left\{\frac{2R}{\kappa^2}+c_{1}R^2+c_{2}R^{\mu\nu}R_{\mu\nu}+\ldots\right\}
\end{equation}
where the ellipses denote that the effective action is in fact an infinite
series---at each new loop order additional higher derivative terms must be
taken
into account. This Lagrangian includes all possible higher derivative
couplings, and
every coupling constant in the Lagrangian is considered to be determined
empirically. Similarly one must include higher derivative contributions to
the
matter Lagrangian in order to treat this piece of the Lagrangian as an
effective
field theory. Details of such considerations can be found in the
papers~\cite{Donoghue:1993eb,Donoghue:dn}.

In our calculations we will consider only the non-analytic
contributions, which are generated by the propagation of two or
more massless particles in the Feynman diagrams. Such nonanalytic
effects are long-ranged and, in the low energy limit of the
effective field theory, they dominate over the analytic
contributions which arise from the propagation of massive modes.
The typical non-analytic terms we will consider are of the type:
$1/\sqrt{-q^2}$ and $\ln q^2$---while the typical analytic
contribution is a power series in $q$. The feature that the
analytic contributions originate from the propagation of massive
particles while the non-analytic effect comes from massless
propagation can be seen directly by Taylor expanding the two types
of propagators. Indeed the massless propagator $1/q^2$ cannot be
expanded in a series, while we have the obvious representation for
the massive propagator: $1/(q^2-m^2) = -1/m^2(1 - q^2/m^2 +
\ldots)$. Thus the massive parts of the diagrams will always be
expandable and hence analytic---while the massless contributions
have the possibility of generating nonanalytic components.  It
should be noted that such nonanalytic pieces of the scattering
amplitude are essential to the unitarity of the S-matrix.

\subsection{Definition of the potential} Before proceeding to the
actual calculations, it is important to note that the precise
definition of a potential in a relativistic quantum field theory
such as general relativity is not obvious. In the original papers,
the one-particle-irreducible potential was
calculated~\cite{Donoghue:1993eb,Donoghue:dn}. However, additional
diagrams are required in order to relate this quantity to physical
observables. Subsequent work has considered alternative
definitions. Clearly a gravitational potential should be gauge
invariant in order to make physical sense, but one can choose
between various definitions of the potential depending on the
physical situation, how the total energy is defined, etc. In QCD
one can always turn to the venerable Wilson loop description but
there exists no standard Wilson loop description for the
gravitational potential, although some work has been done in this
direction~\cite{Modanese:1994bk} using the Arnowitt-Deser-Misner
formula for the total energy of the gravitational system. Such an
approach to the gravitational potential has been taken in the
paper~\cite{Muzinich:1995uj}. For yet another approach to the 
Newtonian potential, see~\cite{Diego}. 
An alternative path is to use the
scattering amplitude itself to define the potential. This
description of the potential seems to us the simplest and most
intuitive picture and has been employed by a number of
authors~\cite{Iwasaki,Okamura,Gupta,Hamber:1995cq,Kazakov:2000mu,ibk}.
Herein then we shall also use the full scattering amplitude in
order to represent the potential, defining
\begin{eqnarray}
\langle f |T| i \rangle &\equiv&
(2\pi)^4\delta^{(4)}(p-p')({\cal M}(q))\nonumber\\
 &=& -(2\pi)\delta(E-E')\langle f| \tilde V({\bf q})|i\rangle
\end{eqnarray}
where $p,p'$ is the incoming, outgoing four-momentum.  The corresponding
coordinate
space representation can be found by taking the nonrelativistic limit and
Fourier-transforming, yielding the result
\begin{equation}
V({\bf x})=\frac{1}{2m_1}\frac{1}{2m_2}\int\frac{d^3q}{(2\pi)^3}e^{i{\bf
q}\cdot{\bf
x}} {\cal M}(\vec{q})
\end{equation}
and will serve as our definition of the nonrelativistic potential.

It should be noted, however, that this is not the only way in
which to define the potential $ V({\bf q})$ in terms of the
scattering amplitude. One could, for example, subtract off the
second order Born contributions, which would lead to the
nonrelativistic potential used in bound state quantum mechanics
and would be equivalent to using the prescription
\begin{eqnarray}
i\langle f |T| i\rangle &=& -2\pi i\delta(E-E')\nonumber\\
&\times&\bigg[\langle f| \tilde V_{bs}({\bf q})| i\rangle +\sum_n
\frac{\langle f
|\tilde V_{bs}({\bf q})| n\rangle \langle n| \tilde V_{bs}({\bf q})|
i\rangle}{E-E_n+i\epsilon} + \ldots\bigg]
\end{eqnarray}
The definition of the bound state potential is discussed in detail in
~\cite{Iwasaki}. In particular, in a Hamiltonian treatment there are also
terms in
the Hamiltonian involving $Gp^2/r$ that contribute at the same order. The
relation
of the bound state potential $\tilde V_{bs}({\bf q})$, in
Einstein-Infeld-Hoffmann
coordinates, to the lowest order scattering potential is
\begin{equation}
\tilde V_{bs} (r)= V(r) + {7 Gm_1m_2 (m_1+m_2)\over 2c^2 r^2}
\end{equation}

\subsection{The diagrams contributing to the nonanalytic component of the
scattering matrix}

We will consider here only the nonanalytic contributions from the
one-loop diagrams. Since many diagrams yield purely analytic
contributions to the S-matrix, such diagrams need not be
considered and will be omitted from the beginning. The diagrams
which {\it do} yield nonanalytic contributions to the $S$ matrix
amplitude are those containing two or more massless propagating
particles.  A typical such amplitude will be of the form
\begin{equation}
{\cal M} \sim
\Big(A+Bq^2+\ldots+\alpha\kappa^4\frac1{q^2}+\beta_1\kappa^4\ln(-q^2)+
\beta_2\kappa^4\frac{m}{\sqrt{-q^2}} +\ldots \Big)
\end{equation}
Here the coefficients $A, B, \ldots$ correspond to analytic pieces
which are of no interest to us, as these terms will only dominate
in the high energy regime of the effective theory. Rather, the
$\alpha, \beta_1, \beta_2, \ldots$ coefficients correspond to the
nonlocal, nonanalytic contributions to the amplitude and are the
ones which we seek. In particular, the $\beta_1, \beta_2$ terms
will yield the leading post-Newtonian and quantum corrections to
the potential.

\section{Results for the Feynman diagrams}
In this section we will present our results. All diagrams have
been performed both by hand and by computer. In order to evaluate
the diagrams by computer, an algorithm for Maple 7
(TM)\footnote{\footnotesize{Maple and Maple V are registered
trademarks of Waterloo Maple Inc.}} was developed. This program
contracts the various indices and perform the loop integrations.
All results obtained this way were confirmed results obtained by
hand. The resulting amplitudes were then Fourier transformed to
produce the scattering potential, and only the nonanalytic pieces
of the amplitude were retained.  For this part of the calculation,
the following Fourier integrals are useful:
\begin{equation}\begin{split}
\int \frac {d^3 q}{(2\pi)^3} ~e^{i{\bf q}\cdot {\bf r}}
\frac 1{|{\bf q}|^2}& = \frac {1}{4\pi r}\\
\int \frac {d^3 q}{(2\pi)^3} ~ e^{i{\bf q}\cdot {\bf r}}
\frac 1{|{\bf q}|}& = \frac {1}{2\pi^2 r^2}\\
\int \frac {d^3 q}{(2\pi)^3} ~e^{i{\bf q}\cdot {\bf r}}
\ln({\bf q}^2)& = \frac {-1}{2\pi r^3}\\
\end{split}\end{equation}
After this brief introduction, we proceed to give the results for
each diagram in turn.  Note that the basic vertices needed for our
calculation are given in Appendix A.

\subsection{The tree diagram}
\begin{figure}[h]\vspace{0.5cm}
\begin{minipage}{\linewidth}
\begin{center}
\includegraphics[scale=1]{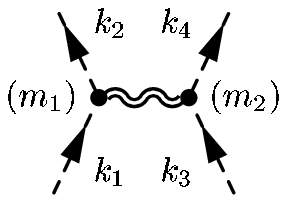}
\end{center}
{\begin{center}1(a)\end{center}}
\caption{The tree diagram giving Newtons law.}\label{tree}
\end{minipage}
\end{figure}
The result for this diagram in the nonrelativistic limit is the well-known
lowest
order tree-level result which yields the Newtonian potential. We define
the diagram
using the Feynman rules as:
\begin{equation}
iM_{1(a)}(\vec{q}) = \tau^{\mu\nu}_1(k_1,k_2,m_1)\Big[\frac{i{\cal
P}_{\mu\nu\alpha\beta}}{q^2}\Big]\tau_1^{\alpha\beta}(k_3,k_4,m_2)
\end{equation}
where $q=k_1-k_2=k_4-k_3$. Contracting all indices and taking the
nonrelativistic
limit, we find
\begin{equation}
M_{1(a)}(\vec{q})=-{4\pi Gm_1m_2\over \vec{q}^2}
\end{equation}
whose Fourier transform produces the scattering potential
\begin{equation}
V_{1(a)}(r) = -\frac{G{m_1}{m_2}}{r}
\end{equation}
which is indeed the familiar Newtonian form.

\subsection{The box and crossed box diagrams}
\begin{figure}[h]\vspace{0.5cm}
\begin{minipage}[t]{0.46\linewidth}
\begin{center}
\includegraphics[scale=1]{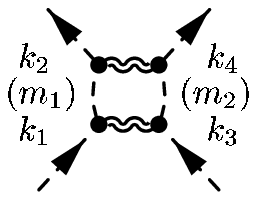}
\end{center}
{\begin{center}{2(a)}\end{center}}
\end{minipage}
\begin{minipage}[t]{0.46\linewidth}
\begin{center}
\includegraphics[scale=1]{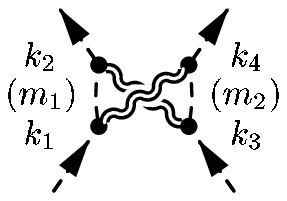}
\end{center}
{\begin{center}2(b)\end{center}}
\end{minipage}
\caption{The box and crossed box diagrams which contribute to the
non-analytic component of the potential.}\label{box}
\end{figure}
We can write the contributions of these two diagrams as
\begin{equation}\begin{aligned}
M_{2a}&=\int\frac{d^4l}{(2\pi)^4}\tau_1^{\mu\nu}(k_1,k_1+l,m_1)
\tau_1^{\rho\sigma}(k_1+l,k_2,m_1)\\ & \times
\tau_1^{\alpha\beta}(k_3,k_3-l,m_2)
\tau_1^{\gamma\delta}(k_3-l,k_4,m_2)\\
& \times
\Big[\frac{i}{(k_1+l)^2-m_1^2}\Big]\Big[\frac{i}{(k_3-l)^2-m_2^2}\Big]\Big[\frac{i{\cal
P}_{\mu\nu\alpha\beta}}{l^2}\Big]\Big[\frac{i{\cal
P}_{\rho\sigma\gamma\delta}}{(l+q)^2}\Big]
\end{aligned}\end{equation}
for the box and
\begin{equation}\begin{aligned}
M_{2b}&=\int\frac{d^4l}{(2\pi)^4}\tau_1^{\mu\nu}(k_1,k_1+l,m_1)
\tau_1^{\rho\sigma}(k_1+l,k_2,m_1)\\ & \times
\tau_1^{\gamma\delta}(k_3,l+k_4,m_2)
\tau_1^{\alpha\beta}(l+k_4,k_4,m_2)\\
& \times \Big[\frac{i}{(k_1+l)^2-m_1^2}\Big]\cdot
\Big[\frac{i}{(l+k_4)^2-m_2^2}\Big]\cdot \Big[\frac{i{\cal
P}_{\mu\nu\alpha\beta}}{l^2}\Big]\Big[\frac{i{\cal
P}_{\rho\sigma\gamma\delta}}{(l+q)^2}\Big]
\end{aligned}\end{equation}
for the cross box.  These diagrams are among the most challenging
that we will encounter, due to the rather complicated
integrals---containing {\it four} propagators---which must be
evaluated. However, those pieces of the amplitude which are loop
momentum-dependent simplify, due to the feature that the external
particles are on shell and by the fact that we are only seeking
the nonanalytic pieces of the scattering amplitude. This allows
reduction of parts of the amplitude initially having four
propagators to pieces where effectively only three or two
propagators remain. An example of this simplification can be seen
by the replacement of the integral
\begin{equation}\begin{aligned}
\int & \frac{d^4l}{(2\pi)^2} \frac{l\cdot
k_1}{l^2(l+q)^2((l+k_1)^2-m_1^2)((l-k_3)^2-m_2^2)}\\ &  =
\frac12\int\frac{d^4l}{(2\pi)^2}
\frac{((l+k_1)^2)- l^2-
m_1^2}{l^2(l+q)^2((l+k_1)^2-m_1^2)((l-k_3)^2-m_2^2)}
\end{aligned}\end{equation}
by
\begin{equation}\begin{aligned}
\frac12\int\frac{d^4l}{(2\pi)^2} \frac{1}{l^2(l+q)^2((l-k_3)^2-m_2^2)}
\end{aligned}\end{equation}
because the $l^2$ part will not contribute to the nonanalytic
component. The
integrals with three or two propagator terms are explicitly given in
Appendix B.
Another simplification arises when the momentum $q^\mu$ contracts with a
loop
momentum $l^\mu$ in the numerator.  An example is
\begin{equation}\begin{aligned}
\int & \frac{d^4l}{(2\pi)^2} \frac{l\cdot
q}{l^2(l+q)^2((l+k_1)^2-m_1^2)((l-k_3)^2-m_2^2)}\\ &  =
\frac12\int\frac{d^4l}{(2\pi)^2}
\frac{((l+q)^2)-l^2-
q_1^2}{l^2(l+q)^2((l+k_1)^2-m_1^2)((l-k_3)^2-m_2^2)}
\end{aligned}\end{equation}
which simplifies to
\begin{equation}\begin{aligned}
-\frac12\int\frac{d^4l}{(2\pi)^2} \frac{q^2}{l^2(l+q)^2((l-k_3)^2-m_2^2)}
\end{aligned}\end{equation}
Via these simplifications one can reduce the box and cross box amplitudes
to a
reduced piece consisting only of integrals with two or three propagators
and a
component with the basic form of the box and crossed box integrals, {\it
i.e.}, with
no loop momentum terms in the numerator.  Then, using the integrals
presented in
Appendix B, performing the above-described contractions in the two
diagrams, and
taking the nonrelativistic limit we end up with the result
\begin{equation}
M_{2(a)+2(b)}^{\rm red}(\vec{q})={46\over 3}G^2m_1m_2\log\vec{q}^2
\end{equation}
for the momentum-reduced component of the box plus cross box and
\begin{equation}
M_{2(a)+2(b)}^{\rm irred}(\vec{q})=16G^2m_1m_2\log\vec{q}^2
\end{equation}
for the irreducible piece.  Fourier transforming, we find then the
scattering
potential contributions
\begin{equation}
V_{2(a)+2(b)}^{\rm red}(r) = -\frac{23}{3}\frac{m_1m_2 G^2}{\pi r^3}
\end{equation}
for the reducible component and
\begin{equation}
V_{2(a)+2(b)}^{\rm irred}(r) = -8\frac{m_1m_2 G^2}{\pi r^3}
\end{equation}
for the irreducible piece so that the total result for the box and cross
box
contribution to the potential is
\begin{equation}
V_{2(a)+2(b)}^{\rm tot}(r) = -\frac{47}{3}\frac{m_1m_2 G^2}{\pi r^3}
\end{equation}
These results are in agreement with those of ~\cite{ibk}.

\subsection{The triangle diagrams}
\begin{figure}[h]
\begin{minipage}[t]{0.46\linewidth}
\begin{center}
\includegraphics[scale=1]{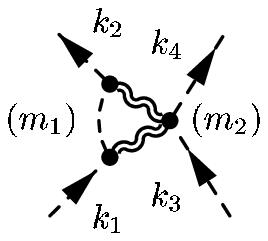}
\end{center}
{\begin{center}3(a)\end{center}}
\end{minipage}
\begin{minipage}[t]{0.46\linewidth}
\begin{center}
\includegraphics[scale=1]{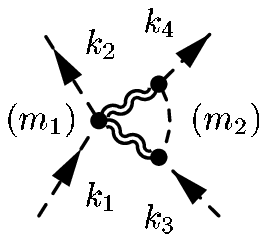}
\end{center}
{\begin{center}3(b)\end{center}}
\end{minipage}
\caption{The set of triangule diagrams contributing to the
scattering potential.}\label{trig}
\end{figure}
The next pieces we will consider are the set of triangle diagrams,
for which we find
\begin{equation}
\begin{aligned}
M_{3(a)}(q)&=\int{d^4l\over (2\pi)^4}\tau_1^{\mu\nu}(k_1,l+k_1,m_1)
\tau_1^{\alpha\beta}(l+k_1,k_2,m_1)\tau_2^{\sigma\rho\gamma\delta}(k_3,k_4,m_2)\\
&\times\Big[\frac{i{\cal P}_{\alpha\beta\gamma\delta}}
        {(l+q)^2}\Big] \Big[\frac{i{\cal P}_{\mu\nu\sigma\rho}}{l^2}\Big]
\Big[\frac{i}{(l+k_1)^2-m_1^2}\Big]
\end{aligned}
\end{equation}
and
\begin{equation}
\begin{aligned}
M_{3(b)}(q)&=\int{d^4l\over (2\pi)^4}\tau_1^{\sigma\rho}(k_3,k_3-l,m_2)
\tau_1^{\gamma\delta}(k_3-l,k_4,m_2)\tau_2^{\mu\nu\alpha\beta}(k_1,k_2,m_1)\\
        &\times\Big[\frac{i{\cal P}_{\mu\nu\sigma\rho}}{l^2}\Big]
        \Big[\frac{i{\cal P}_{\alpha\beta\gamma\delta}}{(l+q)^2}\Big]
\Big[\frac{i}{(l-k_3)^2-m_2^2}\Big]
\end{aligned}
\end{equation}
The calculation of such diagrams yields no real complications--- the
integrals
needed are quite straightforward and are presented in Appendix
B.  However, a
significant simplification results from the use of the identity
\begin{equation}
{\cal P}_{\gamma\delta\sigma\rho}{\cal
P}_{\alpha\beta\mu\nu}\tau^{\sigma\rho\mu\nu}
(k_1,k_2,m_1)=\tau_{\gamma\delta\alpha\beta}(k_1,k_2,m_1)\label{eq:ide}
\end{equation}
and, taking the nonrelativistic limit, we find for these two pieces
\begin{eqnarray}
M_{3(a)}(\vec{q})&=&-8G^2m_1m_2\left({7\over
2}\log\vec{q}^2+{\pi^2m_1\over |\vec{q}|}\right)\nonumber\\
M_{3(b)}(\vec{q})&=&-8G^2m_1m_2\left({7\over
2}\log\vec{q}^2+{\pi^2m_2\over
|\vec{q}|}\right)
\end{eqnarray}
Our results for these diagrams agree with those of ~\cite{ibk}, and the
Fourier
transformed result is
\begin{equation}
V_{3(a)+3(b)}(r) = -4\frac{G^2m_1m_2(m_1+m_2)}{r^2}+28\frac{m_1m_2
G^2}{\pi r^3}
\end{equation}

\subsection{The double-seagull diagram}
\begin{figure}[h]
\begin{center}
\includegraphics[scale=1]{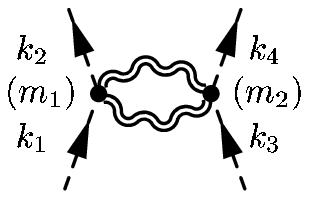}
\end{center}
{\begin{center}4(a)\end{center}} \caption{The double-seagull diagram
contribution to the scattering
potential.}\label{circ}
\end{figure}
We have for the double-seagull term
\begin{equation}
\begin{aligned}
M_{4(a)}(q)&={1\over 2!}\int{d^4l\over (2\pi)^4
}\tau_2^{\alpha\beta\gamma\delta}(k_1,k_2,m_1)\tau_2^{\sigma\rho\mu\nu}(k_3,k_4,m_2)
        \times\Big[\frac{i{\cal P}_{\alpha\beta\mu\nu}}{(l+q)^2}\Big]
        \Big[\frac{i{\cal P}_{\gamma\delta\sigma\rho}}{l^2}\Big]
\end{aligned}
\end{equation}
The double-seagull loop diagram is quite straightforward, and is
simplified by use of the identity Eq. \ref{eq:ide}. Note, however,
that there exists a symmetry factor of $1/2!$.  The resulting
amplitude is found to be
\begin{equation}
M_{4(a)}(\vec{q})=44G^2m_1m_2\log\vec{q}^2
\end{equation}
whose Fourier transform yields the double-seagull contribution to the
potential
\begin{equation}
V_{4(a)}(r) = -22\frac{m_1m_2 G^2}{\pi r^3}
\end{equation}

\subsection{The vertex correction diagrams}
\begin{figure}[h]
\begin{minipage}[t]{0.46\linewidth}
\begin{center}
\includegraphics[scale=1]{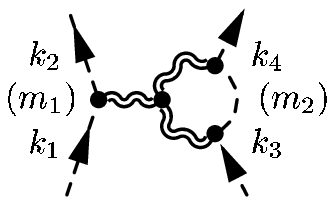}
\end{center}
{\begin{center}5(a)\end{center}}
\end{minipage}
\begin{minipage}{0.1\linewidth}\ \end{minipage}
\begin{minipage}[t]{0.46\linewidth}
\begin{center}
\includegraphics[scale=1.1]{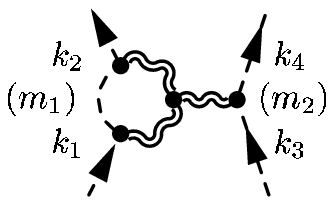}
\end{center}
{\begin{center}5(b)\end{center}}
\end{minipage}
\begin{minipage}[t]{0.46\linewidth}
\begin{center}
\includegraphics[scale=1]{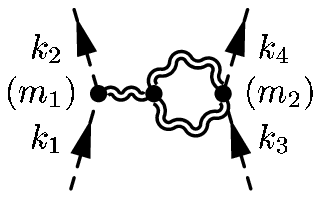}
\end{center}
{\begin{center}5(c)\end{center}}
\end{minipage}
\begin{minipage}{0.1\linewidth}\ \end{minipage}
\begin{minipage}[t]{0.46\linewidth}
\begin{center}
\includegraphics[scale=1]{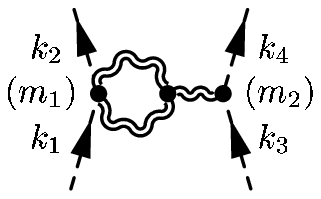}
\end{center}
{\begin{center}5(d)\end{center}}
\end{minipage}
\caption{The class of the graviton vertex corrections which yield
nonanalytic
corrections to the potential.}\label{gravi2n}
\end{figure}
There exist two classes of vertex correction diagrams:  For the
massive loop diagrams, shown in Figures 5(a) and 5(b) we have
\begin{equation}
\begin{aligned}
M_{5(a)}(q)&={d^4l\over (2\pi)^4
}\tau_1^{\alpha\beta}(k_1,k_2,m_1)\tau_1^{\mu\nu}(k_3,k_3-l,m_2)
\tau_1^{\rho\sigma}(k_3-l,k_4,m_2)\tau_3^{\lambda\kappa\phi\epsilon(\gamma\delta)}(l,-q)
\\
 &       \times\Big[\frac{i{\cal P}_{\lambda\kappa\mu\nu}}{l^2}\Big]
 \Big[\frac{i{\cal P}_{\phi\epsilon\rho\sigma}}{(l+q)^2}\Big]
\Big[\frac{i{\cal
P}_{\alpha\beta\gamma\delta}}{q^2}\Big]\Big[\frac{i}{(l-k_3)^2-m_2^2}\Big]
\end{aligned}
\end{equation}
and
\begin{equation}
\begin{aligned}
M_{5(b)}(q)&=\int{d^4l\over (2\pi)^4}\tau_1^{\alpha\beta}(k_1,l+k_1,m_1)
\tau_1^{\mu\nu}(l+k_1,k_2,m_1)\tau_1^{\lambda\kappa}(k_3,k_4,m_2)
\tau_3^{\gamma\delta\rho\sigma(\phi\epsilon)}(-l,q)\\
  &      \times\Big[\frac{i{\cal P}_{\alpha\beta\gamma\delta}}{l^2}\Big]
  \Big[\frac{i{\cal P}_{\mu\nu\rho\sigma}}{(l+q)^2}\Big]
\Big[\frac{i{\cal
P}_{\phi\epsilon\lambda\kappa}}{q^2}\Big]\Big[\frac{i}{(l+k_1)^2-m_1^2}\Big]
\end{aligned}
\end{equation}
while for the pure graviton loop diagrams shown in Figures 5(c)
and 5(d)we have
\begin{equation}
\begin{aligned}
M_{5(c)}(q)&={1\over 2!}\int{d^4l\over (2\pi)^4
}\tau_2^{\lambda\kappa\epsilon\phi}(k_3,k_4,m_2)
\tau_1^{\alpha\beta}(k_1,k_2,m_1)\tau_3^{\mu\nu\rho\sigma(\gamma\delta)}(l,-q)\\
&     \times\Big[\frac{i{\cal P}_{\mu\nu\lambda\kappa}}{l^2}\Big]
   \Big[\frac{i{\cal P}_{\rho\sigma\epsilon\phi}}{(l+q)^2}\Big]
\Big[\frac{i{\cal P}_{\alpha\beta\gamma\delta}}{q^2}\Big]
\end{aligned}
\end{equation}
and
\begin{equation}
\begin{aligned}
M_{5(d)}(q)&={1\over 2!}\int{d^4l\over (2\pi)^4
}\tau_2^{\rho\sigma\mu\nu}(k_1,k_2,m_1)\tau_1^{\lambda\kappa}(k_3,k_4,m_2)
\tau_3^{\alpha\beta\gamma\delta(\epsilon\phi)}(-l,q)\\
   &     \times\Big[\frac{i{\cal P}_{\mu\nu\gamma\delta}}{l^2}\Big]
   \Big[\frac{i{\cal P}_{\rho\sigma\alpha\beta}}{(l+q)^2}\Big]
\Big[\frac{i{\cal P}_{\lambda\kappa\epsilon\phi}}{q^2}\Big]
\end{aligned}
\end{equation}
The vertex correction diagrams are certainly the most challenging
to perform, disregarding the box and crossed box diagrams, and the
results go back to the original calculation
of~\cite{Donoghue:1993eb,Donoghue:dn}---however, due to an
algebraic error, the original result quoted for the such diagrams
was in error.  Since that time, the results for the gravitational
vertex corrections have been checked at length in various
publications~\cite{akh,ibk}; however, until \cite{metric} the
correct forms have not been given.  Using the results of
\cite{metric} and taking the nonrelativistic limit, we find the
amplitudes,
\begin{eqnarray}
M_{5(a)+5(b)}(\vec{q})&=&2G^2m_1m_2\left({\pi^2(m_1+m_2)\over
|\vec{q}|}+{5\over 3}\log\vec{q}^2\right)\nonumber\\
M_{5(c)+5(d)}(\vec{q})&=&-{52\over 3}G^2m_1m_2\log\vec{q}^2
\end{eqnarray}
whose Fourier transform yields the corrected results for the
vertex modifications to the scattering potential:
\begin{equation}
V_{5(a)+5(b)}(r) = \frac{G^2m_1m_2(m_1+m_2)}{r^2}-\frac53\frac{m_1m_2
G^2}{\pi r^3}
\end{equation}
and
\begin{equation}
V_{5(c)+5(d)}(r) = \frac{26}{3}\frac{m_1m_2 G^2}{\pi r^3}
\end{equation}
where again we note the presence of a symmetry factor $1/2!$ in
the case of the pure graviton loop diagram.\footnote{It is
correctly pointed out in \cite{ibk} that the coefficient of the
two-graviton vertex quoted in
\cite{Donoghue:1993eb},\cite{Donoghue:dn} is too small by a factor
of two. However, the numerical result for the loop integrals given
therein is correct because of the presence of this symmetry
factor.}  Our results for these diagrams are {\it not} in
agreement with previous calculations. For the vertex correction
diagrams $5(a)$ and $5(b)$, there is total disagreement with our
result. However, we have performed a very detailed analysis of the
vertex diagrams in our companion paper~\cite{metric}, showing how
the classical ingredients match in detail those required by the
classical Schwartzschild metric.  The quantum vertex result
of~\cite{ibk} is the same as that of \cite{akh} and the latter
yields an incorrect {\it classical} metric (Ref. \cite{ibk} does
not display their classical vertex correction).  On the other
hand, our detailed evaluation of the metric correction
in~\cite{metric} gives us confidence in the correctness of our
result.

\subsection{The vacuum polarization diagram}
\begin{figure}[ht]
\begin{minipage}[t]{0.46\linewidth}
\begin{center}
\includegraphics[scale=1]{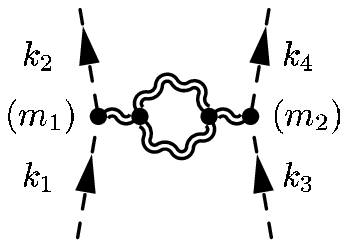}
\end{center}
{\begin{center}6(a)\end{center}}
\end{minipage}
\begin{minipage}{0.1\linewidth}\ \end{minipage}
\begin{minipage}[t]{0.46\linewidth}
\begin{center}
\includegraphics[scale=1]{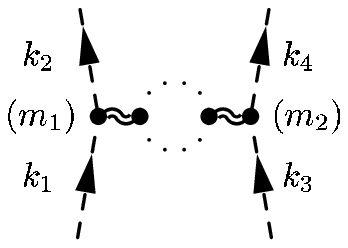}
\end{center}
{\begin{center}6(b)\end{center}}
\end{minipage}
\caption{The vacuum polarization diagrams which contribute to the
potential. Note that there exists a ghost diagram along with the
graviton loop.}\label{vacuum}
\end{figure}
The result for the vacuum polarization contribution can be found
from the amplitude
\begin{equation}
M_{6(a)+6(b)}(q)=\tau_{\rho\sigma}(k_1,k_2,m_1){i{\cal
P}^{\rho\sigma\lambda\xi}\over
q^2}\Pi_{\lambda\xi\mu\nu}(q){i{\cal P}^{\mu\nu\gamma\delta}\over
q^2}\tau_{\gamma\delta}(k_3,k_4,m_2)
\end{equation}
where the vacuum polarization tensor is found from the effective
Lagrangian obtained
by 't Hooft and Veltman~\cite{Veltman1,Veltman2,Goroff:1985th}---
\begin{equation}
{\cal L}=-{1\over 16\pi^2}\log\vec{q^2}\left({1\over 120}R^2+{7\over
20}R_{\mu\nu}R^{\mu\nu}\right)
\end{equation}
and is given by
\begin{eqnarray}
\hat{\Pi}_{\alpha\beta,\gamma\delta}&=&-{2G\over
\pi}\log(-q^2)\left[{21\over
120}q^4 I_{\alpha\beta,\gamma\delta}+{23\over
120}q^4\eta_{\alpha\beta}\eta_{\gamma\delta} \right.\nonumber \\
&-&\left.{23\over 120}q^2(\eta_{\alpha\beta}q_\gamma q_\delta
+\eta_{\gamma\delta}q_\alpha
q_\beta)\right.\\
&-&\left.{21\over 240}q^2(q_\alpha q_\delta\eta_{\beta\gamma}+q_\beta
q_\delta\eta_{\alpha\gamma} +q_\alpha q_\gamma\eta_{\beta\delta}+q_\beta
q_\gamma\eta_{\alpha\delta}) +{11\over 30}q_\alpha q_\beta
q_\gamma q_\delta\right]\nonumber\\
\quad
\end{eqnarray}
Contracting the various indices, we find a result
\begin{equation}
M_{6(a)+6(b)}(\vec{q})={43\over 15}G^2m_1m_2\log\vec{q}^2
\end{equation}
which is equivalent to that originally derived
in~\cite{Donoghue:1993eb,Donoghue:dn}.  After Fourier transforming we find
a
contribution to the scattering potential
\begin{equation}
V_{6(a)+6(b)}(r) = -\frac{43}{30}\frac{m_1m_2 G^2}{\pi r^3}
\end{equation}
which is in agreement with that given by \cite{ibk}.

\section{The result for the gravitational potential}

Adding up all the corrections to the non-relativistic potential we
have our final result
\begin{equation}
V(r) = -{G m_1m_2\over
r}\left[1+3\frac{G(m_1+m_2)}{r}+\frac{41}{10\pi}\frac{
G\hbar}{ r^2}\right]\label{eq:pot}
\end{equation}
The classical term in this potential agrees with Eq. 2.5 of
Iwasaki\cite{Iwasaki}.

However, we note that the quantum component of the potential is not
equivalent to
{\it any} previous published
result~\cite{Donoghue:1993eb,Donoghue:dn,Muzinich:1995uj,Hamber:1995cq,akh,ibk}.
As
described above, we believe this result to be to definitive form of the
nonrelativistic scattering matrix potential. Our results agree with
\cite{ibk} for
all diagrams except the vertex corrections\footnote{To be precise we are
comparing
to the third version of \cite{ibk}, available in the electronic
archive. We had
several disagreements with the original version, all of which except the
vertex
correction have been corrected in the third version.} However we have
multiple
checks on the vertex correction, as we have seen that our version leads
exactly to
the required form of the classical Schwarzschild metric, for both fermions
and
bosons.

\subsection{A Potential Ambiguity}
There is an important issue concerning the potential that has not
been thusfar much discussed in the literature---that both the
classical and quantum corrections to the potential are in some
sense ambiguous. For the classical correction, this realization
goes back to the papers of~\cite{Okamura,Barker:bx, Barker:ae},
and these arguments generalize readily to the quantum component.
In this section we discuss this ambiguity and argue that at one
loop order our calculation of the quantum correction does have a
well defined meaning.

As explored in~\cite{Okamura} the classical post-Newtonian potential is
not
invariant under a coordinate transformation of the form
\begin{equation}
r \rightarrow r\Big[1+\alpha\frac{G(m_1+m_2)}{r}\Big]
~. \label{eqn:class}
\end{equation}
We are always free to make such a coordinate change, and that
given in Eq. \ref{eqn:class} modifies the classical correction to
the potential via
\begin{equation}
\frac{Gm_1m_2}{r} \left[1+c\frac{G(m_1+m_2)}{r}  \right]\rightarrow
\frac{Gm_1m_2}{r} \left[1+(c-\alpha)\frac{G(m_1+m_2)}{r} \right]
\end{equation}
Following~\cite{Okamura,Barker:ae} the ambiguity can be seen to arise in a
field
theory calculation through a modification of the graviton propagator
\begin{equation}
\frac{1}{q^2} = \frac{1}{q_0^2-{\vec q}^2}
\end{equation}
accomplished by using the energy conservation in the following identical
way which
holds true for general $x$.
\begin{equation}
\frac{1}{\frac{1}{2}(1-x)(({p_2}_0-{p_1}_0)^2+({p_4}_0-{p_3}_0)^2))-x({p_2}
_0-{p_1}_0)({p_4}_0-{p_3}_0)-{\vec q}^2}
\end{equation}
Using this propagator in order to derive the gravitational
potential, the result will in general depend on $x$, with the
relation to the coordinate change being $\alpha=-\frac14(1-x)$.
Not only will the terms of order $\frac{G^2}{r^2}$ depend on $x$,
but so will the corrections of order
$\Big(\frac{G}{r}\Big)\Big(\frac{v}{c}\Big)$. This indicates that
the post-Newtonian part of the static gravitational potential is
{\it not} well-defined in and of itself.

The coordinate ambiguity also generalizes to the quantum part of
the potential. There exists a coordinate redefinition
\begin{equation}
r \to r\left[1+ \beta {G\hbar \over r^2}\right]  \label{eqn:quant}
\end{equation}
which changes the coefficient of the quantum term in the potential
\begin{equation}
\frac{Gm_1m_2}{r} \left[1+d\frac{G\hbar}{r^2}  \right]\rightarrow
\frac{Gm_1m_2}{r}
\left[1+(d-\beta)\frac{G\hbar}{r^2} \right]
\end{equation}
Therefore, we need to address the uniqueness of the quantum correction to
the
potential.

Since general relativity is invariant under coordinate shifts, if
the potentials are changed there must exist other modifications
that compensate for these changes, leaving the resulting physics
invariant.  Within a Hamiltonian formulation, these modifications
take the form of momentum-dependent pieces---in order to make the
potential well defined, one must also specify the momentum terms
in the Hamiltonian.

First we examine the classical ambiguity. Classically there are
two dimensionless variables available for the expansion away from
the Newtonian limit
\begin{equation}
{p^2\over m^2}  ~~,~~ {Gm\over r}
\end{equation}
In bound state problems these quantities appear at the same order, as can
be seen by
use of the virial theorem. In the Hamiltonian the leading terms are order
\begin{equation}
H_1 \sim {p^2\over m} ,~~{Gm^2 \over r}
\end{equation}
and are unambiguous. At next order we have pieces of the form
\begin{equation}
H_2 \sim {p^4 \over m^3}, ~~{Gm^2\over r} {Gm\over r}, ~~({Gm\over
r})({p^2\over m})
\end{equation}
There exists an ambiguity between the last two of these terms arising from
a
coordinate change, but this effect cancels for physical observables, as
shown
explicitly in~\cite{Okamura}. Writing the Hamiltonian in the center of
mass frame to
post-Newtonian order, we have
\begin{eqnarray}
H &=& \left(\frac{{\bf p}^2}{2m_1} +\frac{{\bf p}^2}{2m_2}
\right)-\left(
\frac{{\bf p}^4}{8m_1^3} +\frac{{\bf p}^4}{8m_2^3} \right) \nonumber \\
&-&\frac{Gm_1m_2}{r}\left[ 1+a\frac{\bf p^2}{m_1m_2} +b \frac{({\bf p\cdot
\hat{r}})^2}{m_1m_2} + c{G (m_1+m_2)\over r} \right]
\end{eqnarray}
In the standard Einstein-Infeld-Hoffmann coordinates, the coefficients
$a,b,c$ have
the values
\begin{eqnarray}
a &=& \frac12\left[1+3\frac{(m_1+m_2)^2}{m_1m_2}\right]
\nonumber\\
b&=& {1\over 2} \nonumber \\
c &=& -\frac12
\end{eqnarray}
but under the coordinate transformation of Eq. \ref{eqn:class}, one finds
the
modifications
\begin{eqnarray}
a &\to&\frac12\left[1+(3+2\alpha)\frac{(m_1+m_2)^2}{m_1m_2}\right]
\nonumber\\
b &\to&  \frac12-\alpha \frac{(m_1+m_2)^2}{m_1m_2} \nonumber \\
c &\to& - \frac12 - \alpha
\end{eqnarray}
Therefore, in order to specify the correction to the static potential, one
needs
also to identify the momentum coordinates. In particular, the classical
potential
calculated by Iwasaki is that appropriate for Einstein-Infeld-Hoffmann
coordinates.

This classical ambiguity has not been worked out explicitly at the
following order
in the expansion, but the general pattern is apparent.  Specifically, at
the next
order one has four types of terms
\begin{equation}
H_3 \sim \frac{p^6}{m^5},~~ \frac{p^4}{m^3}\left({Gm\over r}\right),
~\frac{p^2}{m}\left({Gm\over r}\right)^2,~~{Gm^2\over r}\left({Gm\over
r}\right)^2
.
\end{equation}
The last term here is noteworthy because it goes like $1/r^3$---{\it i.e.}
like the
quantum correction in the potential---but it is distinguishable by its
mass
dependence plus the fact that it is order $G^3$.  Such terms will also be
ambiguous
under a coordinate transformation, but the effects will cancel among the
effects
terms of this order. So the``classical'' coordinate change above does
change the
$1/r^3$ term in the potential, but has a specific form and the effects
cancels among
other classical terms in the Hamiltonian.

Now let us examine the quantum effects, keeping only one power of the
quantum
expansion parameter
\begin{equation}
{G\hbar \over r^2}
\end{equation}
The quantum potential will enter the Hamiltonian at order
\begin{equation}
H_q = {Gm^2\over r}{G\hbar\over r^2}
\end{equation}
and the other term of this order is
\begin{equation}
H_{qp} \sim {p^2 \over m} {G\hbar \over r^2}
\end{equation}
If one makes the``quantum'' coordinate change of Eq. \ref{eqn:quant} ,
this will
generate such a term in the Hamiltonian, which for physical observables
cancels the
effect of the change in the potential. Therefore in order to make the
quantum
correction to the potential well defined, one must specify the value of
the terms
contained in $H_{qp}$. However, the important point is that, in {\bf{any}}
coordinates in which one calculates quantum corrections, one will {\it
not} generate
a term such as $H_{qp}$. This is because all quantum effects that involve
the
interparticle separation $r$ arise from loop diagrams of order $G^2$ while
$H_{qp}$
has only a single power of $G$. Quantum effects on a single particle line
are of
order $G$ but do not involve $r$, while those involving two lines are of
order
$G^2$. Since all calculations yield no term of the form $H_{qp}$, they
must give the
same quantum potential. The fully specified quantum potential is that one
in the
coordinates where $H_{qp}$ vanishes. This, of course, is then the one that
we have
calculated. We conclude that, despite the expected general coordinate
invariance,
the calculated quantum potential is a well defined quantity to this order.

\section{Discussion}

The scattering amplitude has been used to provide a solid
definition of the quantum corrections to the Newtonian potential.
Our basic calculation is of certain nonanalytic terms in momentum
space, the $\sqrt{-q^2}$ and $q^2 \ln -q^2$ terms.  A
transformation of these terms to coordinate space allows us to
interpret these as long distance corrections to the potential. Our
result is displayed in Eq. \ref{eq:pot}.

The quantum corrections that we find for the scattering potential
hold equally for the bound state potential. This is because the
transformation between the two has no quantum component. The
difference between the two is a classical correction that comes
from iterating the lowest order potential. Dimensional analysis
reveals that in the nonrelativistic limit this iteration has the
ability to generate only a classical effect and not a quantum
correction.

We have found a result for the nonrelativistic potential which we
believe is the final and complete result for this quantity. The
potential matches the expectations from dimensional analysis as
discussed previously~\cite{Donoghue:1993eb,Donoghue:dn} and the
known ambiguity of the form of the classical correction has been
seen to originate from the possibility of rewriting a potential
energy term in the Hamiltonian in terms of kinetic energy and vice
versa, as also discussed in~\cite{Okamura,Barker:ae,Barker:bx}.
Such rewritings are not possible to do for the quantum term at the
order to which we work.  Therefore the quantum correction to the
potential is a definite exact quantity.

The quantum corrections are too small to be observed
experimentally. However, the
fact that these are reliably predicted is important for our understanding
of quantum
gravity. These effects are due to the low energy propagation of massless
degrees of
freedom and hence are uniquely predicted for any quantum theory of gravity
that
reduces to general relativity in the low energy limit\footnote{Indeed,
related
effects have been found within the context of M-theory \cite{becker}
although the
precise coefficients have not been calculated}. In this sense, these are
low energy
theorems of quantum gravity.

\begin{center}
{\bf Acknowledgments}
\end{center}
N.E.J. Bjerrum-Bohr would like to thank the Department of Physics and
Astronomy at
UCLA for its kind hospitality and P.H. Damgaard for discussions.  The work
of BRH
an JFD is supported in part by the National Science Foundation under award
PHY-98-01875.

\begin{center}
{\Large\bf Appendices}
\end{center}
\appendix

\section{Vertices and Propagators}
We begin by listing the Feynman rules which are employed in our
calculation. For a
derivation of these forms, see \cite{metric}.

\subsection{Scalar propagator}\noindent
The massive scalar propagator is:\\ \\

\begin{minipage}[h]{0.4\linewidth}\vspace{0.4cm}
\centering\includegraphics[scale=1]{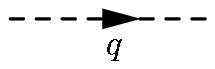}
\end{minipage}
\begin{minipage}[h]{0.65\linewidth}
\centering$\displaystyle = \frac i{q^2-m^2+i\epsilon}$
\end{minipage}

\subsection{Graviton propagator}\noindent The graviton propagator
in harmonic gauge can be written in the form:\\ \\

\begin{minipage}[h]{0.4\linewidth}\vspace{0.4cm}
\centering\includegraphics[scale=1]{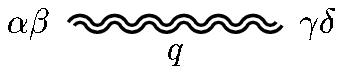}
\end{minipage}
\begin{minipage}[h]{0.65\linewidth}
\centering $\displaystyle = \frac {i{\cal
P}^{\alpha\beta\gamma\delta}}{q^2+i\epsilon}$
\end{minipage}\\
where $${\cal P}^{\alpha\beta\gamma\delta} =
\frac12\left[\eta^{\alpha\gamma}\eta^{\beta\delta} +
\eta^{\beta\gamma}\eta^{\alpha\delta}
-\eta^{\alpha\beta}\eta^{\gamma\delta}\right]$$

\subsection{2-scalar-1-graviton vertex}\noindent The
2-scalar-1-graviton vertex is discussed in the literature. We
write it as:\\

\begin{minipage}[h]{0.4\linewidth}\vspace{0.15cm}
\centering\includegraphics[scale=1]{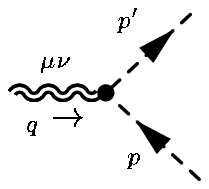}
\end{minipage}
\begin{minipage}[h]{0.65\linewidth}
\centering$\displaystyle = \tau_1^{\mu \nu}(p,p',m)$
\end{minipage}\vspace{0.3cm}
where
$$\tau_1^{\mu\nu}(p,p',m) = -\frac{i\kappa}2\left[p^\mu p^{\prime \nu}
+p^\nu p^{\prime \mu} - \eta^{\mu\nu}\left((p\cdot
p^\prime)-m^2\right)\right]
$$

\subsection{2-scalar-2-graviton vertex}
The 2-scalar-2-graviton vertex is also discussed in the
literature. We write it here with the full symmetry of the two
gravitons:\\

\begin{minipage}[h]{0.4\linewidth}\vspace{0.15cm}
\centering\includegraphics[scale=1]{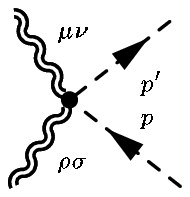}
\end{minipage}
\begin{minipage}[h]{0.65\linewidth}
\centering$\displaystyle = \tau_2^{\eta\lambda\rho\sigma}(p,p',m)$
\end{minipage}\vspace{0.3cm}
where
\begin{equation}{\begin{aligned}
\tau_2^{\eta \lambda \rho \sigma}(p,p')&= {i\kappa^2} \bigg [ \left
\{I^{\eta
\lambda\alpha \delta} {I}^{\rho \sigma\beta}_{\ \ \ \ \delta} -
\frac14\left\{\eta^{\eta \lambda} I^{\rho \sigma\alpha \beta} +
 \eta^{\rho \sigma} I^{\eta \lambda\alpha \beta} \right \}
\right \} \left (p_\alpha p^\prime_{\beta} + p^\prime_{\alpha} p_\beta
\right ) \\
&-\frac12 \left \{ I^{\eta \lambda\rho \sigma} - \frac12\eta^{\eta
\lambda}\eta^{\rho \sigma} \right \} \left [ (p\cdot p') - m^2
\right]\bigg]\end{aligned}}
\end{equation}
with
$$I_{\alpha\beta\gamma\delta}={1\over
2}(\eta_{\alpha\gamma}\eta_{\beta\delta}
+\eta_{\alpha\delta}\eta_{\beta\gamma}).$$
\subsection{3-graviton vertex}
The 3-graviton vertex can be derived via the background field method and
has the
form\cite{Donoghue:1993eb},\cite{Donoghue:dn}

\begin{minipage}[h]{0.4\linewidth}\vspace{0.15cm}
\centering\includegraphics[scale=1]{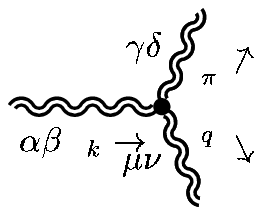}
\end{minipage}
\begin{minipage}[h]{0.65\linewidth}
\centering$\displaystyle =
{\tau_3}_{\alpha\beta\gamma\delta}^{\mu\nu}(k,q)$
\end{minipage}\vspace{0.3cm}
where
\begin{equation}{
\begin{aligned}
{\tau_3}_{\alpha  \beta \gamma \delta }^{\mu
\nu}(k,q)&=-\frac{i\kappa}2\times
\bigg({\cal P}_{\alpha \beta \gamma \delta }\bigg[k^\mu k^\nu+ (k-q)^\mu
(k-q)^\nu
+q^\mu q^\nu-
\frac32\eta^{\mu \nu}q^2\bigg]\\[0.00cm]&
+2q_\lambda q_\sigma\bigg[ I_{\alpha \beta }^{\ \ \
\sigma\lambda}I_{\gamma \delta
}^{\ \ \ \mu \nu} + I_{\gamma \delta }^{\ \ \ \sigma\lambda}I_{\alpha
\beta }^{\ \ \
\mu \nu} -I_{\alpha \beta }^{\ \ \ \mu  \sigma} I_{\gamma \delta }^{\ \ \
\nu
\lambda} - I_{\gamma \delta }^{\ \ \ \mu \sigma} I_{\alpha \beta }^{\ \ \
\nu
\lambda}
\bigg]\\[0cm]&
+\bigg[q_\lambda q^\mu \bigg(\eta_{\alpha \beta }I_{\gamma \delta }^{\ \ \
\nu
\lambda}+\eta_{\gamma \delta }I_{\alpha \beta }^{\ \ \ \nu
\lambda}\bigg) +q_\lambda
q^\nu \left(\eta_{\alpha \beta }I_{\gamma \delta }^{\ \ \ \mu
\lambda}+\eta_{\gamma
\delta }I_{\alpha \beta }^{\ \ \ \mu  \lambda}\right)\\&
-q^2\left(\eta_{\alpha
\beta }I_{\gamma \delta }^{\ \ \ \mu \nu}-\eta_{\gamma \delta }I_{\alpha
\beta }^{\
\ \ \mu \nu}\right) -\eta^{\mu \nu}q_\sigma q_\lambda\left(\eta_{\alpha
\beta
}I_{\gamma \delta }^{\ \ \ \sigma\lambda} +\eta_{\gamma \delta }I_{\alpha
\beta }^{\
\ \
\sigma\lambda}\right)\bigg]\\[0cm]&
+\bigg[2q_\lambda\big(I_{\alpha \beta }^{\ \ \ \lambda\sigma}I_{\gamma
\delta
\sigma}^{\ \ \ \ \nu}(k-q)^\mu +I_{\alpha \beta }^{\ \ \
\lambda\sigma}I_{\gamma
\delta \sigma}^{\ \ \ \ \mu }(k-q)^\nu -I_{\gamma \delta }^{\ \ \
\lambda\sigma}I_{\alpha \beta \sigma}^{\ \ \ \ \nu}k^\mu -I_{\gamma \delta
}^{\ \ \
\lambda\sigma}I_{\alpha \beta \sigma}^{\ \ \ \ \mu }k^\nu \big)\\&
+q^2\left(I_{\alpha \beta \sigma}^{\ \ \ \ \mu }I_{\gamma \delta }^{\ \ \
\nu
\sigma} + I_{\alpha \beta }^{\ \ \ \nu \sigma}I_{\gamma \delta \sigma}^{\
\ \ \ \mu
}\right) +\eta^{\mu \nu}q_\sigma q_\lambda\left(I_{\alpha \beta }^{\ \ \
\lambda\rho}I_{\gamma \delta  \rho}^{\ \ \ \ \sigma} +I_{\gamma \delta
}^{\ \ \
\lambda\rho}I_{\alpha \beta  \rho}^{\ \ \ \
\sigma}\right)\bigg]\\[0cm]&
+\bigg\{(k^2+(k-q)^2)\big[I_{\alpha \beta }^{\ \ \ \mu  \sigma}I_{\gamma
\delta
\sigma}^{\ \ \ \ \nu} +I_{\gamma \delta }^{\ \ \ \mu  \sigma}I_{\alpha
\beta
\sigma}^{\ \ \ \ \nu} -\frac12\eta^{\mu \nu}{\cal P}_{\alpha \beta \gamma
\delta
}\big]\\&-\left(I_{\gamma \delta }^{\ \ \ \mu \nu}\eta_{\alpha \beta
}k^2+I_{\alpha
\beta }^{\ \ \ \mu \nu}\eta_{\gamma \delta }(k-q)^2\right)\bigg\}\bigg)
\end{aligned}}
\end{equation}

\section{Useful Integrals}
In the evaluation of the various diagrams we employ the following
integrals
\begin{eqnarray}\displaystyle
J=&\displaystyle \int\frac{d^4l}{(2\pi)^4} \frac{1}{l^2(l+q)^2} & =
\frac{i}{32\pi^2}\big[-2L\big] + \ldots\\
J_\mu=&\displaystyle\int\frac{d^4l}{(2\pi)^4} \frac{l_\mu}{l^2(l+q)^2} & =
\frac{i}{32\pi^2}\Big[q_\mu L\Big]+\ldots\\
J_{\mu\nu}=&\displaystyle\int\frac{d^4l}{(2\pi)^4} \frac{l_\mu
l_\nu}{l^2(l+q)^2}
 & = \frac{i}{32\pi^2}\bigg[q_\mu q_\nu \Big(-\frac23L\Big)-
q^2\eta_{\mu\nu}\Big(-\frac16 L\Big)\bigg]+\ldots
\end{eqnarray}
as well as
\begin{eqnarray}\displaystyle
I&=&\displaystyle\int\frac{d^4l}{(2\pi)^4}
\frac{1}{l^2(l+q)^2((l+k)^2-m^2)}
\nonumber\\ & =&
\frac{i}{32\pi^2m^2}\big[-L-S\big]+\ldots\\
I_\mu&=&\displaystyle\int\frac{d^4l}{(2\pi)^4}
\frac{l_\mu}{l^2(l+q)^2((l+k)^2-m^2)}
\nonumber\\ & =& \frac{i}{32\pi^2m^2}\bigg[k_\mu\bigg(\Big(-1-\frac12
\frac{q^2}{m^2}\Big)L-
\frac14\frac{q^2}{m^2}S\bigg)+q_\mu\bigg(L+\frac12S\bigg)\bigg]+\ldots\\
I_{\mu\nu}&=&\displaystyle\int\frac{d^4l}{(2\pi)^4} \frac{l_\mu
l_\nu}{l^2(l+q)^2((l+k)^2-m^2)} \nonumber\\ & =&
\frac{i}{32\pi^2m^2}\bigg[q_\mu
q_\nu\bigg(-L-\frac38 S\bigg) +k_\mu k_\nu
\bigg(-\frac12\frac{q^2}{m^2}L-\frac18\frac{q^2}{m^2}S\bigg)\nonumber\\
&+&\big(q_\mu k_\nu + q_\nu k_\mu \big)\bigg(\Big(\frac12 +
\frac12\frac{q^2}{m^2}\Big)L + \frac{3}{16}\frac{q^2}{m^2}S\bigg)+
q^2\eta_{\mu\nu}\Big(\frac14L+\frac18S\Big)
\bigg]+\ldots\nonumber\\
\quad
\end{eqnarray}
\begin{eqnarray}\displaystyle
I_{\mu\nu\alpha}&=&\displaystyle\int\frac{d^4l}{(2\pi)^4}
\frac{l_\mu l_\nu l_\alpha}{l^2(l+q)^2((l+k)^2-m^2)} \nonumber\\
&=& \frac{i}{32\pi^2m^2}\bigg[ q_\mu q_\nu
q_\alpha\bigg(L+\frac5{16}S\bigg)+k_\mu
k_\nu
k_\alpha\bigg(-\frac16 \frac{q^2}{m^2}\bigg) \nonumber\\
\nonumber&+&\big(q_\mu k_\nu k_\alpha + q_\nu k_\mu k_\alpha + q_\alpha
k_\mu
k_\nu\big)\bigg(\frac13\frac{q^2}{m^2}L+
\frac1{16}\frac{q^2}{m^2}S\bigg)\nonumber\\&+&\big(q_\mu q_\nu k_\alpha +
q_\mu
q_\alpha k_\nu + q_\nu q_\alpha k_\mu \big)\bigg(\Big(-\frac13 -
\frac12\frac{q^2}{m^2}\Big)L
-\frac{5}{32}\frac{q^2}{m^2}S\bigg)\nonumber\\
\nonumber &+&\big(\eta_{\mu\nu}k_\alpha + \eta_{\mu\alpha}k_\nu +
\eta_{\nu\alpha}k_\mu\big)\Big(\frac1{12}q^2L\Big)\nonumber\\
\nonumber&+&\big(\eta_{\mu\nu}q_\alpha + \eta_{\mu\alpha}q_\nu +
\eta_{\nu\alpha}q_\mu\big)\Big(-\frac16q^2L -\frac1{16}q^2S\Big)
\bigg]+\ldots\nonumber\\
\quad
\end{eqnarray}
where we have defined $L=\ln(-q^2)$ and
$S=\frac{\pi^2m}{\sqrt{-q^2}}$. Only the lowest order nonanalytic
terms are included in the above forms. Higher order nonanalytic
contributions as well as the neglected analytic terms are denoted
by the ellipses. The following identities can be verified on
shell, $k\cdot q = \frac{q^2}2$, where $k-k'=q$ and
$k^2=m^2=k^{\prime 2}$. In some cases the integrals are used with
$k$ replaced by $-k'$, where $k' \cdot q = -\frac{q^2}{2}$.

In order to do evaluate the box diagrams the following lowest
order integrals are needed. The higher order contributions of
non-analytic terms as well as neglected analytic terms are again
denoted by ellipses.
\begin{eqnarray}\displaystyle
K  &= &\displaystyle \int\frac{d^4l}{(2\pi)^4}
\frac{1}{l^2(l+q)^2((l+k_1)^2-m_1^2)((l-k_3)^2-m_2^2)}\nonumber\\
&= & \frac{i}{16\pi^2m_1 m_2
q^2}\bigg[\Big(1-\frac{w}{3m_1m_2}\Big)L\bigg]+\ldots \\
K' &= &\displaystyle \int\frac{d^4l}{(2\pi)^4}
\frac{1}{l^2(l+q)^2((l+k_1)^2-m_1^2)((l+k_4)^2-m_2^2)}\nonumber\\
&= & \frac{i}{16\pi^2m_1 m_2
q^2}\bigg[\Big(-1+\frac{W}{3m_1m_2}\Big)L\bigg]+\ldots
\end{eqnarray}
Here $k_1\cdot q = \frac{q^2}2$, $k_2\cdot q = -\frac{q^2}2$,
$k_3\cdot q = -\frac{q^2}2$ and $k_4\cdot q = \frac{q^2}2$, where
$k_1-k_2=k_4-k_3=q$ and $k_1^2=m_1^2=k_2^2$ together with
$k_3^2=m_2^2=k_4^2$. Also, we have defined $w = (k_1\cdot
k_3)-m_1m_2$ and $W = (k_1\cdot k_4) - m_1m_2$.

The following constraints for the nonanalytic terms of the above
integrals hold true on shell:
\begin{equation}
I_{\mu\nu\alpha}\eta^{\alpha\beta} = I_{\mu\nu}\eta^{\mu\nu} =
J_{\mu\nu}\eta^{\mu\nu} = 0
\end{equation}
\begin{equation} \begin{split}
I_{\mu\nu\alpha}q^\alpha = -\frac{q^2}{2}I_{\mu\nu}, \ \ I_{\mu\nu}q^\nu
= -\frac{q^2}{2}I_{\mu}, \ \ I_{\mu}q^\mu  = -\frac{q^2}{2}I \ \ \\
J_{\mu\nu}q^\nu = -\frac{q^2}{2}J_{\mu}, \ \  J_{\mu}q^\mu   =
-\frac{q^2}{2}J
\end{split} \end{equation}
\begin{equation}
I_{\mu\nu\alpha}k^\alpha = \frac{1}{2}J_{\mu\nu}, \ \ \ \ I_{\mu\nu}k^\nu
=
\frac{1}{2}J_{\mu}, \ \ \ \   I_{\mu}k^\mu  = \frac{1}{2}J
\end{equation}

\end{document}